\documentclass[twocolumn,showpacs,preprintnumbers,amsmath,amssymb,superscriptaddress]{revtex4}
\usepackage{graphicx}
\usepackage{dcolumn}
 \usepackage{bm}
\usepackage{verbatim}   
\usepackage{color}      
\usepackage{subfigure}  
\usepackage{hyperref}   
\usepackage{pstricks}
\usepackage{epstopdf}
\usepackage{eucal}
\usepackage[english]{babel}
\usepackage{amsmath,amsfonts,amssymb}
\usepackage{latexsym}
\usepackage[dvips]{epsfig}

\begin{document}


\title{ ELECTRON PAIRING  AND COULOMB REPULSION \\
IN ONE-DIMENSIONAL ANHARMONIC LATTICES }

\author{ L. Brizhik}
\affiliation{
Bogolyubov Institute for Theoretical Physics, Metrolohichna Str., 14b, Kyiv 03680, Ukraine}
\affiliation{
Instituto Pluridisciplinar, Universidad Complutense, Paseo Juan XXIII, 1, Madrid 28040, Spain}
\affiliation{Wessex Institute of Technology, Ashurst, Southampton SO40 7AA, UK}
\author{ A.P. Chetverikov}
\affiliation{
Instituto Pluridisciplinar, Universidad Complutense, Paseo Juan XXIII, 1, Madrid 28040, Spain}
\affiliation{
Faculty of Physics, Chernyshevsky State University, Astrakhanskaya b. 83, Saratov 410012, Russia}
\author{ W. Ebeling}
\affiliation{
Instituto Pluridisciplinar, Universidad Complutense, Paseo Juan XXIII, 1, Madrid 28040, Spain}
\affiliation{
Institut f\"ur Physik, Humboldt Universit\"at, Newtonstrasse 15, Berlin 12489, Germany}
\author{ G. R\"opke}
\affiliation{
Instituto Pluridisciplinar, Universidad Complutense, Paseo Juan XXIII, 1, Madrid 28040, Spain}
\affiliation{
 Institut f\"ur Physik, Universit\"at Rostock, Rostock 18051, Germany}
\author{M.G. Velarde}
\affiliation{
Instituto Pluridisciplinar, Universidad Complutense, Paseo Juan XXIII, 1, Madrid 28040, Spain}
\affiliation{Wessex Institute of Technology, Ashurst, Southampton SO40 7AA, UK}

\date{\today}

\begin{abstract}
We show that in anharmonic one-dimensional crystal lattices pairing of electrons or holes in a localized \textit{bisolectron} state is possible due to coupling between the charges and the lattice deformation that can overcompensate the Coulomb repulsion. Such localized soliton-like states appear as traveling ground bound \textit{singlet}  states of two extra electrons in the potential well created by the  local lattice deformation. We also find the first excited localized state of two electrons in a soliton-like lattice deformation potential well given by a \textit{triplet} state of two electrons. The results of the analytical study of interacting electrons in a lattice with \textit{cubic} anharmonicity are compared with the numerical simulations of two electrons in an  anharmonic lattice 
taking into account of the (local) Hubbard electron-electron repulsion.
We qualitative agreement between both approaches for a broad interval of parameter values. For illustration we give expressions for the bisolectron binding energy with parameter values that are typical for biological macromolecules. We estimate critical values of Coulomb repulsion where the bisolectron becomes unbound.
\end{abstract}

\maketitle

\section{\label{sec:introduction}
 Introduction}

A broad class of low-dimensional systems such as polydiacetylene
\cite{polydiacet1,polydiacet2,Donovan,polydiacet3}, conducting platinum chain compounds and conducting polymers \cite{Monceau}, salts of transition metals (PbSe, PbTe, PbS) \cite{trans-met1,trans-met2,trans-met3,trans-met4},
superconducting cuprates \cite{cuprates1,cuprates2,cuprates3,cuprates4,8,7} which find numerous applications in microelectronics and nanotechnologies, or play important role in living  systems (polypeptide macromolecules, DNA, etc.) 
\cite{ASD,Scott,Chris,Bountis,Dauxois,Yakush,Peyrard,Manevich}, manifest nonlinear effects. These effects arise from the electron-lattice interaction or from the (nonlinear) lattice \textit{anharmonicity} and, eventually, from both items coupled together. As a consequence there is the formation of soliton-like states, that have been called electro-solitons or solectrons in different albeit related contexts \cite{ASD,16}. Both  generalize the polaron concept \cite{Landau,Pekar,Kaganov,5,Alex,Alex2,6}. It has been shown that \textit{anharmonic} lattices can support the formation of two-component solitons \cite{ASD-Zolot1,ASD-Zolot3,ASD-Zolot2} and pairing of two excess electrons or holes \cite{Briz-Dav,13,Briz-Er,1,2}
 may be enhanced bringing strong correlation both in \textit{momentum} space and in \textit{real} space.
 
 The problem for pairing of charged particles is the Coulomb repulsion. In normal metallic superconductors, the attractive interaction needed to form a bound state of electrons (Cooper pair) is due to phonon excitations in the \textit{harmonic} lattice. Lattice polarization overcomes the Coulomb repulsion so that for states near to the Fermi surface, where the repulsion is rather weak, pairing becomes possible (in \textit{momemtum} space only with complete delocalization in real space). This situation changes in the strong coupling case where the two-electron bound state is localized in space on one lattice site (small bipolaron) \cite{5,Alex}. 
 
For one-dimensional electron-lattice systems with \textit{cubic} or \textit{quartic} anharmonicity,  we have shown \cite{1,2} that excess electrons with  opposite spins may form a \textit{bisolectron}, which is a localized bound state of the paired electrons in a soliton-like lattice  deformation. 
When one \textit{excess} electron is localized on several lattice sites forming a \textit{solectron} state, that is a bound electron-soliton state, the Coulomb repulsion between two electrons in a bisolectron is not as strong as in a small bipolaron, but is not negligible as in the case of completely delocalized electrons.

In contrast to two- and three-dimensional systems, the Coulomb interaction can lead to divergent potential energy in one-dimensional systems if both electrons are found with finite density at the same point in space. Nevertheless, this singularity can be removed by using the Hubbard model of a discrete lattice \cite{3,4}. We consider in Sec. II the case when the electron-lattice interaction is moderately strong, so that an adiabatic approximation is valid which leads to the soliton-like solutions of the system of coupled nonlinear differential equations in the continuum approximation. Then we study first the case when the Coulomb interaction is relatively weak thus modifying weakly the two-electron bisolectron wave function (Sec. III) and, subsequently, the rather strong Coulomb repulsion case for which we find the two-electron wave function within a variational approach (Sec. IV). Solutions for the ground state and the first excited state are considered in Sec. V. We compare these analytical results with results of numerical simulations for two excess electrons in the lattice in Sec. VI. 
The lattice interaction between the nearest neighbors is described by the Morse potential that gives also a  cubic anharmonicity. The electrons are considered within a Hubbard model.  
Finally, in Section VII we draw some conclusions, discussing the applicability of the continuum model and the stability of the bisolectron state.

\section{Dynamics of the coupled electron-lattice system}
\label{Sec:2}

We consider an infinitely long, one-dimensional (1d) crystal lattice, with units all of equal mass $M$ and equilibrium lattice spacing $a$, where two free, excess electrons are added. The Hamiltonian of such a system can be represented in the form:
\begin{equation}
\hat{H} = \hat{H}_{\rm el}+\hat{H}_{\rm lat}+\hat{H}_{\rm int}+\hat{H}_{\rm Coul}.
\label{Ham}
\end{equation}

Let $E_0$ denote the on-site electron energy, let $J$ denote the electron exchange interaction energy, and let $\hat{c}^{\dag}_{n,s}$ and $\hat{c}_{n,s}$ be the creation and annihilation operators of an electron with the spin projection $s=\uparrow,\downarrow$ at the site $n$.  Then the electron Hamiltonian in (\ref{Ham}) has the explicit form
\begin{equation}
 \hat{H}_{\rm el}=\sum_{n,s}\left[E_0 \hat{c}^{\dag}_{n,s} \hat{c}_{n,s}-J\hat{c}^{\dag}_{n,s} \left(\hat{c}_{n+1,s}+\hat{c}_{n-1,s}\right)\right].
 \label{Hamel}
\end{equation}
With the account of only longitudinal displacements of atoms from their equilibrium positions, the lattice part of the Hamiltonian
 in (\ref{Ham}) has the form
\begin{equation}
 \hat{H}_{\rm lat}=\sum_{n}\left[\frac{\hat p^2_n}{2 M}+\hat U_{\rm lat} \left(\hat u_n\right)\right],
\label{Hamlat}
\end{equation}
where $\hat u_n $ is an operator of the $n$-th atom displacement, $\hat p_n $ is the operator of the canonically conjugated momentum,
$\hat U_{\rm lat}$ is the operator of the lattice potential energy, whose properties will be defined below.

We consider the case when the dependence of the on-site electron energy on atom displacements is much stronger than that of the exchange interaction energy, so that the electron-lattice interaction part of the Hamiltonian (\ref{Ham}) has the form
\begin{equation}
\hat{H}_{\rm int}=\chi \sum_{n,s}\left(\hat u_{n+1}-\hat u_{n-1}\right)
  \hat{c}^{\dag}_{n,s} \hat{c}_{n,s},
 \label{Hamint}
\end{equation}
where $\chi $ is the electron-lattice interaction constant.

The Coulomb repulsion between the electrons is given by the Hamiltonian
\begin{equation}
 \hat{H}_{\rm Coul}=\sum_{n,m,s,s'}U_{nm} \hat{c}^{\dag}_{n,s}\hat{c}^{\dag}_{m,s'} \hat{c}_{m,s'} \hat{c}_{n,s},
 \label{HamCoul}
\end{equation}
where $ U_{nm} $ is the matrix element corresponding to the Coulomb interaction  of electrons on sites $n$ and $m$.

In the Born-Oppenheimer approximation we can factorize the state vector of the system into two parts,
\begin{eqnarray}
 |\Psi(t) \rangle= |\Psi_{\rm el}(t) \rangle |\Psi_{\rm lat}(t) \rangle.
 \label{BO}
\end{eqnarray}
Here the state vector of the lattice has the form of the product of the operator of coherent displacements of the atoms and the vacuum state of the lattice
\begin{eqnarray}
 |\Psi_{\rm lat}(t) \rangle&=& S(t)  |0 \rangle _{\rm lat}, \nonumber \\
 S(t)&=& \exp \left\{ - \frac{i}{\hbar} \sum_n \left[u_n(t)  \hat p_n- p_n(t) \hat u_n \right] \right\},
 \label{lat}
\end{eqnarray}
where  $u_n(t),\,p_n(t)$   are, respectively, the mean values of the displacements of atoms from their equilibrium positions and their canonically conjugated momenta in the state (\ref{BO}).

In the case of two excess electrons with spins $s_1,s_2$ in the lattice, the electron state-vector has the form
\begin{eqnarray}
  |\Psi_{\rm el}(t) \rangle=\sum_{n_1,n_2,s_1,s_2} \Psi(n_1,n_2,s_1,s_2;t)\hat{c}^{\dag}_{n_1,s_1} \hat{c}^{\dag}_{n_2,s_2}|0\rangle_{\rm el},
 \label{elec}
\end{eqnarray}
where the function $\Psi(n_1,n_2,s_1,s_2;t)$ satisfies the normalization condition
\begin{equation}
\sum_{n_1,n_2,s_1,s_2}|\Psi(n_1,n_2,s_1,s_2;t)|^2=1.
\label{normal1}
\end{equation}

In the absence of a magnetic field the electron wave function of the system can be represented by the product of the spatial and the spin wave functions. The antisymmetry requirement of the two-electron wave function may be fulfilled by the symmetry of the spatial wave function and the antisymmetry of the spin function (singlet state) or by the antisymmetry of the coordinate function and the symmetry of the spin wave function (triplet state).
Using the state vector (\ref{BO})-(\ref{elec}) we can calculate the Hamiltonian functional $H$, corresponding to the Hamiltonian operator (\ref{Ham}):
\begin{eqnarray}
 H=\langle \Psi(t) |\hat{H}|\Psi(t) \rangle .
 \label{aveH}
\end{eqnarray}
Minimizing the functional (\ref{aveH}) with respect to electron and phonon variables, we derive the system of two coupled equations for the two-electron wave function and the atom displacements. In the continuum approximation $n \rightarrow x \equiv na$ this system of equations has the form:
\begin{eqnarray}
&&i \hbar \frac{\partial \Psi(x_1,x_2,t)}{\partial t}=-\frac{\hbar^2}{2 m} \left(\frac{\partial^2 }{\partial x_1^2}
 +\frac{\partial^2 }{\partial x_2^2}\right)\Psi(x_1,x_2,t)
\nonumber
\\
&&+
\chi a \left(\frac{\partial u(x,t)}{\partial x}|_{x=x_1}
 +\frac{\partial u(x,t)}{\partial x}|_{x=x_2}\right)  \Psi(x_1,x_2,t)\nonumber\\&&
+ U_{\rm Coul}(x_1,x_2) \Psi(x_1,x_2,t), \label{SEel1}
\end{eqnarray}
 \begin{eqnarray}
&& \frac{\partial^2 u}{\partial t^2}-V^2_{\rm ac} \frac{\partial^2 U_{\rm lat}}{\partial \rho^2} \frac{\partial^2 u}{\partial x^2}
 \nonumber \\&&
 = \frac{a}{M}\chi \left(\frac{\partial }{\partial x_1}\int dx_2  |\Psi(x_1,x_2,t)|^2|_{x_1=x}\right. \nonumber \\ && \left.
 + \frac{\partial }{\partial x_2}\int dx_1 |\Psi(x_1,x_2,t)|^2|_{x_2=x} \right).
  \label{SElat1}
\end{eqnarray}
Here, $V_{\rm ac}=a \sqrt{w/M}$ is the sound velocity of the lattice, $w$ is the lattice elasticity coefficient, $\Psi(x_1,x_2,t)$ is the two-electron spatial wave function with $x_i=n_i a$, $i=1,2$, $u (x,t)$ is the function of the lattice displacement, and
$U_{\rm Coul}$ accounts for the Coulomb repulsion between the two electrons. Note that to study the case of strong electron localization 
we have to include into  equation (\ref{SElat1}) an additional term $\propto \partial^4 u / \partial x^4$  which takes into account the dispersion of the lattice and can lead to the supersonic solutions, see, e.g., \cite{16a}.

Introducing the function of the lattice deformation
\begin{equation}
\rho(x,t)= -\frac{1}{a} \frac{\partial u(x,t)}{\partial x},
\label{deform}
\end{equation}
 we can write the potential energy of the system appearing in the right hand side of the Schr\"odinger equation  (\ref{SEel1}) in the form
\begin{eqnarray}
U_{\rm tot}=U_{\rm sol}(x_1,x_2,t)+ U_{\rm Coul}(x_1,x_2),
 \\
\label{SEel2}
U_{\rm sol}(x_1,x_2,t)= - \chi a \left[\rho(x_1,t)+\rho(x_2,t)\right],
\label{Usol}
\end{eqnarray}
where the function $U_{\rm sol}(x_1,x_2,t)$ plays the role of an effective potential created by the deformation of the lattice (index 'sol' indicates that this potential is a soliton-like, as it will be shown below).

For the case of the lattice with cubic anharmonicity the potential energy of the lattice, measured in units of $J$, is
\begin{equation}
U_{\rm lat} (\rho)=\frac{1}{2}\rho^2+\frac{\gamma }{3} \rho^3,
\label{lat-potential}
\end{equation}
where $\gamma$ is the anharmonicity parameter whose value is positive, as we wish to focus on rather strong compressions for which the repulsion part of the potential is what really matters.

The solution of Eqs. (\ref{SEel1})-(\ref{SElat1}) without account of the Coulomb repulsion, i.e., when in Eq. (\ref{SEel1}) $U_{\rm Coul}$ is omitted, was found previously \cite{1} for the singlet two-electron problem. In particular, it was found that the lattice deformation is given by the expression
\begin{equation}
\rho  (\xi )= \rho _0 {\rm sech} ^2 (\kappa \xi ), \quad \xi = (x-Vt)/a,
\label{deform-sol}
\end{equation}
with
\begin{equation}
\kappa = \frac{1}{2} \sqrt{\sigma \rho_0}, \qquad \sigma = \frac{\chi a}{J}.
\label{kappa}
\end{equation}
The maximum value of the lattice deformation $\rho_0 $ depends on the soliton velocity $V$ and lattice anharmonicity. For the stationary case, $V=0$, it is given approximately by the analytical expression
\begin{equation}
\rho_0(V=0)  \approx \frac{1}{2}\alpha^2 \gamma^2 \left[1+\frac{3}{144}\alpha^2 \gamma^3 \right],
\label{rho00}
\end{equation}
with
\begin{equation}
 \alpha = \frac{2\delta}{\gamma} \sqrt{\sigma}, \quad \delta=\frac{\chi a}{MV^2_{\rm ac}}=\frac{\chi}{aw}.
\label{param}
\end{equation}
For arbitrary values $V$ of the velocity, $\rho_0(V) $ can be found numerically.

The solution for the spin-singlet two-electron wave function is
\begin{equation}
\Psi(x_1,x_2,t)=\frac{1}{\sqrt{2}}\left[\Psi_1(x_1,t) \Psi_2(x_2,t)+ \Psi_2(x_1,t) \Psi_1(x_2,t)\right].
\label{2el-product}
\end{equation}
The single-electron wave functions $\Psi_1(x,t) =\Psi_2(x,t)\equiv \Psi (x,t) $ are normalized to 1, according to Eq. (\ref{normal1}). The modulus $|\Psi (x,t)|\equiv  \Phi(\xi)$ is found  to be of the soliton-type form
\begin{equation}
 \Phi(\xi)=\sqrt{\frac{\rho_0}{2 \delta}} {\rm sech} (\kappa \xi) \sqrt{1-s^2+\gamma \rho_0  {\rm sech}^2 (\kappa \xi)}
 \label{comsol}
\end{equation}
with $s^2=V^2/V^2_{\rm ac}.$ The phase factor contains the energy and will no be given here. Thus, the parameter $ \kappa$ in Eqs. (\ref{deform-sol}) and (\ref{comsol}) determines the width of electron localization in the bisolectron state (\ref{comsol}), formed due to binding of the two electrons with the lattice deformation in the soliton state (\ref{deform-sol}) without account of the Coulomb repulsion. As shown in \cite{1},  the singlet bisolectron is always a bound state compared with two independent solectrons if the Coulomb repulsion is neglected. The following relation for the binding energy has been found:
\begin{eqnarray}
\label{Ebind}
&&E^{(\rm bind)}= 2 E^{(\rm sol)}(V)-E^{(\rm bis)}(V)\\&&
=\chi a \left[\rho_0 \frac{\frac{4}{3} \gamma \rho_0+1-s^2}{ \gamma \rho_0+1-s^2}-
\rho_0^{(\rm sol)} \frac{\frac{4}{3} \gamma \rho_0^{(\rm sol)}+1-s^2}{ \gamma \rho_0^{(\rm sol)}+1-s^2}\right] >0\,.\nonumber
\end{eqnarray}
Here, $E^{(\rm sol)}$ and $\rho_0^{(\rm sol)}$ are the energy and maximum lattice deformation for the solectron state where one electron is bound to the lattice soliton. The corresponding values are given in Ref. \cite{1}. For instance, typical values for
macromolecules like polypeptides (see \cite{Scott}) are: $\chi =35 - 62 $ pN, $a= 5.4 \cdot 10^{-10} $ m, $w = 39 - 58$ N/m, $M= 5.7 \cdot 10^{-25}$ kg, $V_{\rm ac}= (3.6 - 4.5) \cdot 10^3$ m/s. The transfer energy value for the amid-I excitation is $J_{\rm exc}=
0.001$ eV, and the the electron transfer energy is of the order $J \approx 0.1 - 0.5 $ eV.  For these parameter values, the bisolectron binding energy (\ref{Ebind}) in macromolecules can be estimated in the region $E^{(\rm bind)}  \approx 0.05 - 0.5$ eV, if the Coulomb repulsion is neglected.

\section{Limit case of weak Coulomb repulsion }

If the bisolectron state is extended over few lattice sites, i.e., if the width of the bisolectron $l^{(\rm bis)}a=2\pi a/\kappa $ is bigger than $a$, the Coulomb repulsion is relatively weak as comparing with the binding energy of the bisolectron (\ref{comsol}). Therefore, the wave function (\ref{comsol}) can be generalized as (see \cite{Briz-Dav,13,2})
 \begin{eqnarray}
&&  \Phi_{1,2}(\xi)=\sqrt{\frac{\rho_0}{2 \delta}} \\&& \times {\rm sech} [\kappa (\xi\mp l/2)] \sqrt{1-s^2+\gamma \rho_0  {\rm sech}^2  [\kappa (\xi\mp l/2)] }\,.
\nonumber
 \label{comsol2}
\end{eqnarray}
Here the parameter $l$ accounts for the Coulomb repulsion in Eq.(\ref{SEel1}), so that approximately
\begin{equation}
U_{\rm Coul}=\frac{e^2}{4\pi \epsilon la},
\label{VCoul-sol}
\end{equation}
and determines the distance between the centre of mass (c.o.m.) coordinates of one-electron wave functions. Here $\epsilon$ denotes the dielectric constant of the lattice. Note that here and below we use SI units, and $\epsilon =\epsilon_{\rm rel} \epsilon_0$ contains the relative dielectric constant $\epsilon_{\rm rel}$ of the medium. 

The parameter $l$ can be determined from the condition of the minimum of energy. In the weak Coulomb repulsion case where $2 \pi/\kappa \gg a$, the total energy of the system is
\begin{eqnarray} \label{ebis}
&& E^{(\rm bis)}=2 E_0+\frac{2}{3} J \kappa^2 \frac{\rho_0}{\delta}- \frac{4}{3}  \frac{\chi a \rho^2_0}{\kappa \delta} \left(1-l^2 \kappa^2\right) \\&&
+w a^2  \rho^2_0 \left[ \frac{2}{3}+ \frac{1}{2}\gamma \rho_0^2 -l^2 \kappa^2\left(\frac{1}{3}
+\frac{1}{2} \gamma \rho_0^2 \right)\right]+\frac{e^2}{4\pi \epsilon l a}\,.\nonumber
\end{eqnarray}
Minimizing this expression with respect to $l$, we get the equilibrium distance between the maxima of one-electron functions:
 \begin{equation}
l_0=\frac{1}{2}  \left( \frac{e^2}{\pi \epsilon a \zeta} \right)^{1/3}
 \label{lnull}
\end{equation}
where 
 \begin{eqnarray}
&&\zeta= \left[ \frac{4}{3}  \frac{\chi a \rho^2_0\kappa}{ \delta} -w a^2  \rho^2_0 \kappa^2 \left( \frac{1}{3}+ \frac{1}{2}\gamma \rho_0^2\right)\right]\,.
 \label{zeta}
\end{eqnarray}

Then (\ref{lnull}) can be approximated by 
 \begin{eqnarray}
&& l_0=\frac{1}{2}  \left( \frac{3 \delta e^2}{4 \pi \epsilon \chi a^2  \rho^2_0 \kappa} \right)^{1/3}\,.
 \label{lnull1}
\end{eqnarray}
We remind here that this approach strictly speaking, is valid for relatively weak Coulomb repulsion, so that the condition $l_0\ll l^{(\rm bis)}=2 \pi/\kappa$ is fulfilled.
Thus we find a lattice soliton binding two electrons together \cite{1} in the lattice deformation potential well (\ref{SEel2}).
The bisolectron solution is stable when its binding energy is larger than the Coulomb repulsion that means with Eq. (\ref{Ebind}) $J g^2/2> U_{\rm Coul}$, where $g=\chi^2/(2 J w) = (0.9 - 3.9)$ for polypeptides \cite{Scott}.
Therefore we conclude that the critical value of the electron-lattice coupling constant increases with increasing Coulomb repulsion that agrees with the results of the numerical simulations given in Ref. \cite{dias3}. Note that the Coulomb interaction is screened by the relative dielectric constant $\epsilon_{\rm rel}$ of the order of 10.
For the parameter values given above, we find that $l_0$ is small compared with $2 \pi/\kappa$ so that the weak coupling condition holds. The estimation for the binding energy $ E^{(\rm bis)}$ gives a value of about 0.05 eV indicating a stable bisolectron solution.

In Section VI, Fig \ref{fig5b}, we will show the charge density distribution, $q(\xi)=\Phi_1^2(\xi)+ \Phi_2^2(\xi)$ (see Eq. (\ref{comsol})) for three different values of $l_0$ (see (\ref{lnull}) or (\ref{lnull1})), in other words, for three different relations between the binding energy of the bisolectron and Coulomb repulsion  (\ref{VCoul-sol}). We will see there that the density profile has one maximum at relatively small values of $l_0$ or two maxima at large values of $l_0$.

These results, as will be shown in Section VI, with high degree of accuracy explain the results of numerical simulations of two electrons in a discrete Morse lattice with account of Hubbard interaction between the electrons in a broad interval of the strength of interaction.

\section{Separation of the two-electron center of mass motion }

Here we consider the case of arbitrary Coulomb repulsion, when the Coulomb term in Eq. (\ref{SEel1}) can be represented as
\begin{equation}
U_{\rm Coul}=\frac{e^2}{4\pi \epsilon |x_1-x_2|}.
\label{VCoul-strong}
\end{equation}
Let us introduce now the c.o.m. coordinate $X=(x_1+x_2)/2-Vt$, the running wave coordinate of the soliton
moving with velocity $V$,  and the relative coordinate  $x=x_2-x_1$.
The interaction of electrons with the lattice is described by the effective deformational potential (\ref{Usol}), that, according to the results found in the previous Section, is created by the soliton and may bind the two electrons together. We are assuming now that the solution for the soliton potential $U_{\rm sol}(x_1,x_2,t)$ can be represented in the form:
 \begin{eqnarray}
&& U_{\rm sol}(x_1,x_2,t) \approx U^{(0)}_{\rm sol}(x_1,x_2,t)\nonumber \\&&
= - \chi a \left[\rho^{(0)}(x_1,t)+\rho^{(0)}(x_2,t)\right]
\nonumber \\&& \approx - 2 \chi a \rho^{(0)}(X,t)+U_{\rm rel}(x,t;X),
 \label{solpot}
\end{eqnarray}
 where superscript ${(0)}$ indicates solutions without account of the Coulomb interaction, and where we expand $U_{\rm rel}(x,t;X)$  to second order with respect to $x$.
We use the approximation
 \begin{eqnarray} \label{relpot}
&&  U_{\rm rel}(x,t;X) \approx
\frac{1}{2} m \omega_0^2 x^2,\\&&
m \omega_0^2=- \frac{1}{2}  \chi a \frac{\partial^2\rho^{(0)}(X,t)}{\partial X^2}\Big|_{X=0}=\frac{\chi \rho_0 \kappa^2}{a}.\nonumber
\end{eqnarray}
 Since the potential (\ref{solpot}) splits into two parts we may separate the Hamiltonian functional
\begin{eqnarray}
&& H_{\rm eff}=-\frac{\hbar^2}{4 m}\frac{\partial^2}{\partial X^2} -\frac{\hbar^2}{ m}\frac{\partial^2}{\partial x^2}
\nonumber \\ && -2  \chi a \rho^{(0)}(X,t)
+\frac{1}{2} m \omega_0^2 x^2+\frac{e^2}{4\pi \epsilon |x|}\, .
 \label{Hameff}
\end{eqnarray}
Here we have used only the first non-vanishing term of the expansion of the potential around its minimum.  Higher orders can be treated by perturbation theory. In general, the relative potential well in the co-moving coordinates $U_{\rm rel}(x,t;X)$ depends not only on the relative coordinate  $x$ but also on the center of mass position $X$ so that the relative motion and the c.o.m. motion become coupled.

Our aim was to separate the problem into a c.o.m. problem identical to the previous task without Coulomb repulsion and a parabolic problem for the c.o.m. motion. This way we have first to solve the soliton problem for the common motion to find the function $\rho^{(0)}(X,t)$ and $\omega_0$, and then to solve the parabolic problem including the Coulomb term (\ref{VCoul-strong}).

With the ansatz $\Psi(x_1,x_2,t)= \Phi(X) \phi(x)$,  $E = E_{\rm c.o.m.} + E_{\rm rel}$,  we get the equation for the center of mass motion
 \begin{equation}
\left[ -\frac{\hbar^2}{4 m a^2}\frac{d^2}{d \xi^2} -2 \chi a \rho^{(0)}(\xi)\right] \Phi(a \xi)=E_{\rm c.o.m.} \Phi(a \xi).
 \label{compot}
\end{equation}
where $\xi=X/a$.

\begin{figure}
\includegraphics[width=5cm,angle=-90]{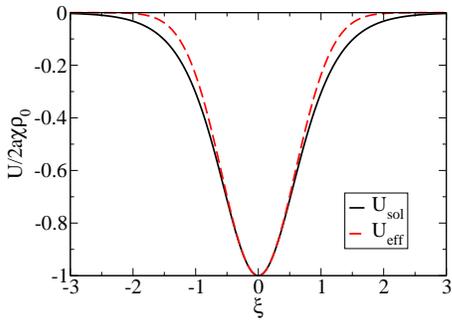}
\caption{
Deformation potential (\ref{comsol3})  for 
$\gamma=1.5,\, \delta=0.002$, $\rho_0=0.007$ and $\kappa=1.2,$ (black solid line) and a Gaussian  (\ref{Gauss})
fitted to the region around the minimum
(red dashed line). Potentials are given in units $2\chi a \rho_0$.
}\label{fig}
\end{figure}

Here the soliton deformation $\rho^{(0)}(\xi)$ is determined in Eq. (\ref{deform-sol}), and, therefore the soliton part of the deformation potential 
(\ref{solpot}) is given by expression
\begin{equation}
U_{\rm sol}(\xi)=- 2 \chi a \rho_0  {\rm sech}^2 (\kappa \xi).
 \label{comsol3}
\end{equation}
We can approximate expression (\ref{comsol3}) by the following one
\begin{equation}
U_{\rm eff}(\xi )=- 2 \chi a \rho_0    e^{-  \kappa^2\xi^2},
\label{Gauss}
\end{equation}
that coincides with expression (\ref{comsol3}) up to the order $\xi^2$. For illustration in Fig. \ref{fig} we show the deformation potentials for the case $\gamma = 1.5$ and other parameters corresponding to their values in poly-peptide macromolecules. For the given parameter values the deformation potentials given by expressions (\ref{comsol3})  and (\ref{Gauss}) show negligible difference, especially near the minimum, which determines the potential $U_{\rm rel}$
(\ref{relpot}) of relative motion. The parabolic approximation (\ref{relpot}) is possible only for $x \le a/\kappa$. For larger distances, the deformation potential becomes weak.

\section{Wave function for the two-electron relative motion
including Coulomb repulsion}

Let us now solve the Schr\"odinger  equation for the relative motion of electrons:
\begin{equation}
\left( -\frac{\hbar^2}{ m}\frac{d^2}{d x^2}+\frac{m \omega_0^2}{2}  x^2+\frac{e^2}{4 \pi \epsilon |x|}\right)
 \phi(x)= E_{\rm rel} \phi(x)
 \label{Hamrel3}
\end{equation}
with the symmetry condition for fermions that in the spin singlet state $\phi(-x)=\phi(x)$ (symmetric orbital) and in the spin triplet state $\phi(-x)=-\phi(x)$ (antisymmetric orbital).

The minimum of the potential follows from (\ref{Hamrel3}) at $x=x_0$,  where
\begin{eqnarray}
&& x_0= \left(\frac{e^2}{4 \pi \epsilon m \omega_0^2}\right)^{1/3}
\label{l0}
\end{eqnarray}
that may be considered as a characteristic length scale. In a classical treatment, it gives the distance between the two electrons at zero temperature.

In the absence of Coulomb repulsion, the solutions for the harmonic potential are Hermitean functions. The ground state is given by the symmetric orbital that has no nodes, but has a finite density at $x=0$ that leads to a divergent energy if Coulomb repulsion is considered. Therefore, with account of Coulomb repulsion we have in the 1d case the additional condition $\phi(0)=0$ for strongly localized electrons, i.e., only solutions with a node at $x=0$ are possible.
To consider the behavior near $x=0$, we have $\phi(x) \propto x+e^2 m/(2 \hbar^2 \epsilon)x^2$. Therefore, both electrons are kept apart by a distance of the order $x_0$ given by the value (\ref{l0}). Comparing this expression with the result (\ref{lnull}), it should be mentioned that now the 
deformation of the lattice remains fixed when the distance between the electrons changes. 

A rigorous solution can be given in the free case $\omega_0=0$ using hypergeometric functions
\begin{eqnarray}
&&
 \phi(x)= {\rm const} \, x \exp\left[-i (m E_{\rm rel}/\hbar^2)^{1/2}x\right]
 \\&&\times {}_1F_1\left[1-
 i \frac{e^2}{4 \pi \epsilon (E_{\rm rel} \hbar^2/m)^{1/2}},2,2i(m E_{\rm rel}/\hbar^2)^{1/2}x\right] \,.\nonumber
 \label{phifree}
\end{eqnarray}
This solution approximates the full solution with the harmonic potential (or another potential that is finite near $x=0$).
Including the harmonic potential, the full solutions can be found numerically. As a peculiarity of the one-dimensional Coulomb problem, symmetric solutions $\phi(-x)= \phi(x)$ and antisymmetric solutions $\phi(-x)= -\phi(x)$ become degenerated. This is an artifact that arises performing the continuum limit for strongly localized solutions and is removed if we consider the discrete lattice and take into account the higher dimensionality of real systems (even so-called one-dimensional systems are in fact confined three-dimensional systems).

In \cite{1} we discuss the solution $\phi(x)$ of the Schr\"odinger equation for the relative motion using a variational approach.
Taking into account the properties we already discussed, we consider the normalized function
 \begin{eqnarray}
&&
 \phi_1(x;\beta)= 2 \left(\frac{2 \beta^3}{\pi}\right)^{1/4} x e^{- \beta x^2}
 \label{phivar}
\end{eqnarray}
that is the Hermitean function for the first excited state. We can improve this ansatz using, e.g., higher Hermitean functions or other zero-node functions like $x\,  {\rm sech} (\beta_1 x)$. The parameter $\beta$ is now a free variational quantity. In the variational problem the energy is the sum of the kinetic energy, harmonic potential and the Coulomb part.
For the class of functions (\ref{phivar}) we have
 \begin{eqnarray}
&&
 E_1(\beta)= 3\frac{\hbar^2}{m}\beta+\frac{3 m \omega_0^2}{8 \beta} + \sqrt{\frac{8}{\pi}}\frac{e^2}{4 \pi \epsilon} \beta^{1/2}\,.
 \label{Erelvar}
\end{eqnarray}
The condition for the minimum gives the minimal energy $E_1(\beta_1)$ and the corresponding
lengths $\beta_1^{-1/2}$ according to
 \begin{eqnarray}
&&
 3\frac{\hbar^2}{m}-\frac{3 m \omega_0^2}{8 \beta_1^2} + \sqrt{\frac{2}{\pi}}\frac{e^2}{4 \pi \epsilon} \beta_1^{-1/2}=0\,.
 \label{Erelmin}
\end{eqnarray}
In the limit where the Coulomb term can be neglected, we find $\beta_1=m \omega_0/(2 \sqrt{2} \hbar)$.
In the opposite case where the Coulomb part is strong, we have
$\beta_1=\left( \sqrt{\frac{\pi}{2}}\frac{3 m \omega_0^2}{8} \frac{4 \pi \epsilon}{e^2}\right)^{2/3},$
so that $\beta_1^{-1/2}$ is proportional to $x_0$, Eq. (\ref{l0}). As an interpolating formula we can take
 \begin{eqnarray}
&&
\beta_1^{-2}= \frac{8 \hbar^2}{m^2 \omega_0^2}+\left( \sqrt{\frac{\pi}{2}}\frac{3 m \omega_0^2}{8} \frac{4 \pi \epsilon}{e^2}\right)^{-4/3}\,.
 \label{Erelminint}
\end{eqnarray}

A more sophisticated ansatz for the variational treatment would be the following class of functions
 \begin{eqnarray}
&&
 \phi^{\rm as}_2(x;\beta,l)=  \left(\frac{ \beta}{2 \pi}\right)^{1/4}\frac{1}{\sqrt{1-e^{-2 \beta l^2}}}
 \nonumber\\&&\times \left[e^{- \beta (x-l)^2}
 -e^{- \beta (x+l)^2}\right]\,.
 \label{phivar2}
\end{eqnarray}
In the limit $l \to 0$ the Hermitean function (\ref{phivar}) is recovered.
The corresponding energy functional $E^{\rm as}_2(\beta,l)=\langle \phi^{\rm as}_2 |H_{\rm rel}|\phi^{\rm as}_2 \rangle$ for the Hamiltonian of the relative motion  can also be given in analytical form that is very tedious and will  be discarded here.

With respect to the symmetric solution, we can take $|\phi_1(x;\beta)|$ from Eq. (\ref{phivar}) that gives the same
energy value as the antisymmetric one. A symmetric wave function $ \phi^{\rm s}_2(x;\beta,l)$ that has no nodes can be obtained as symmetric superposition similar to Eq. (\ref{phivar2}). However, the energy is diverging because $ \phi^{\rm s}_2(0;\beta,l) \ne 0$. If the artificial singularity of the Coulomb repulsion is removed considering the distribution of the electron wave function in the space as given by the atomic orbits, the symmetric solution may become favorable.
For instance, we can replace the Coulomb repulsion by $e^2/(4 \pi \epsilon \sqrt{x^2+a^2})$ where $a$ is of the order of the Bohr radius. Then, the degeneration between the symmetric and antisymmetric solution will be removed so that the symmetric solution (spin singlet) becomes energetically favorable. We will discuss this point below in context with the Hubbard model.

%
%
\begin{figure}
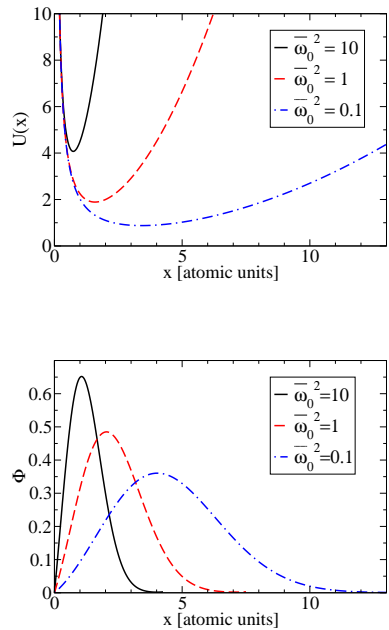

\begin{center} 
\includegraphics[width=5cm]{U.eps}\\
\vspace{1cm}
\includegraphics[width=5cm]{phi.eps}
\end{center}
\caption{\label{fig1}
Potential energy of the relative motion (in Rydberg units) and wave function for the relative motion of the electrons from the numerical solution of the Schr\"odinger equation (\ref{Hamrel3}), for different parameter values of $\bar \omega_0^2=m\omega_0^2/$Ryd.}
\end{figure}

Instead of the variational approach, the Schr\"odinger equation (\ref{Hamrel3}) for the relative motion can be solved numerically.
Potentials for different parameter values of $\bar \omega_0$ (in Rydberg units, $\hbar = e^2/(8 \pi \epsilon) = 2 m=1$) are shown in Fig.~\ref{fig1}. The corresponding normalized wave functions are also shown in Fig.~\ref{fig1} for $x> 0$. For $x < 0$ we have the symmetric solution $\phi(-x)=\phi(x)$ and the antisymmetric solution $\phi(-x)=-\phi(x)$ that are degenerated.

We see that the variational ansatz (\ref{phivar}) can be used and coincides well with the exact solution. However, we point out that the harmonic approximation for the deformation potential is justified only near the minimum of the potential, i.e. $x\le a/\kappa$.

To consider also larger value of the relative distance $x$, we can approximate the interaction by a potential of solitonic form, instead using the harmonic potential, 
\begin{equation}
\left( -\frac{\hbar^2}{m a^2} \frac{d^2}{d \bar x^2}- \bar u {\rm sech}^2(\bar \kappa \bar x)+\frac{e^2}{4 \pi \epsilon a |\bar x|} \right)
 \phi(\bar x)=  E_{\rm rel} \phi(\bar x).
 \label{Hamrel4}
\end{equation}
The relative distance $\bar x=x/a$ is given in units of $a$. The solution of this equation indicates the range of parameter values where a bisolectron solution exists. With the parameter values given for polypeptides \cite{Scott}, we 
estimate the values $\hbar^2/m a^2\approx 0.2$ eV, $e^2/4 \pi \epsilon a  \approx 0.5$ eV. A bound state exists if $\bar u$ exceeds a critical value $\bar u_c$. As example, we find $\bar u_c=0.4742$ eV for $\bar \kappa=0.3$,   $\bar u_c=0.1368$ eV for $\bar \kappa=0.1$. We can estimate $\bar \kappa \approx \kappa/\sqrt{2}, \bar u \approx \chi a \rho_0$. If we consider the value for $E^{(\rm bind)}$ given at the end of Sec. \ref{Sec:2} to characterize the parameter range of $\bar u$, a bound bisolectron state is stable for  values of $\bar \kappa$ below the corresponding critical values that are in the region of parameter values relevant for polypeptides. Note that our approach, which considers the relative motion of the two electrons separately from the center-of-mass motion, works also in the Hubbard models for the  bisolectrons considered in \cite{20,23,23a}. There, the potential strength $-2a \chi \rho_0$ is replaced by $-U_e$ that is the minimum of the polarization potential.



\section{Comparison with numerical results for the Hubbard model }
\label{sec6}

Our treatment of bisolectron states in one-dimensional anharmonic lattices contains two basic approximations. To obtain analytical expressions, we considered the discrete lattice in continuum approximation. Because the extension of the bisolectron state is not very large, the discrete structure of the lattice may become of relevance. In particular, it is not simple to relate the parameter values of the continuum model with the properties that describe the lattice. Thus, the Coulomb repulsion is overestimated a short distances in the continuum model, and the singularity that occurs at zero distance between the two electrons is an artifact. This singularity is removed when the Hubbard model is introduced. 
The other approximations that can be controlled are the expansions in deriving the nonlinear potential and the separation of the center of mass motion from the relative motion. For the parameter values considered here, these approximations
seem to be reasonable and can be improved adding further terms of the expansions.

We will compare now our analytical results with simulations based on the Hubbard Hamiltonian, coupled to a discrete lattice with Morse interaction \cite{20,22} (for related work see \cite{Neissner,18,dias,dias2,dias3}).
The Hamiltonian we use for the simulations describes a 1d periodic lattice  of molecules in which two excess
electrons have been injected.
For estimation we give some typical parameter values of the Morse potential
\begin{equation}
U^{\rm Morse}(r) = D[(1-e^{-B (r-a)})^2-1] \label{46}
\end{equation}
that are adapted to describe biomolecules:
the lattice spacing $a \simeq 1 - 5$ \AA , the stiffness 
$B a \simeq 1$, the depth 
$D \simeq 0.1 - 0.5$ eV. The period of oscillations in the Morse well
results as $1 / \Omega_{\rm Morse} \simeq 0.1 - 0.5 $ ps. 

Noteworthy is that for the Toda potential \cite{30}, with about the same repulsive component, we know the exact analytical solutions that, for the Morse potential (\ref{46}), can be used to quite a satisfactory level of approximation \cite{16,31,32}.
This potential energy of  a lattice with Morse interaction can be expanded in a power series with respect to the displacements of the  atoms from equilibrium positions. The terms beyond the quadratic one describe the anharmonicity of the lattice. Here, we are interested in the cubic term that can be compared with our analytical approach.

The electron part is given by the 1d Hubbard Hamiltonian, cf. Eqs. (\ref{Hamel}), (\ref{HamCoul}),
\begin{eqnarray}
H_{\rm el}+H_{\rm Coul}&=&-\sum_{n,s}\, J_{n,n+1}\left({\hat{c}}_{n+1 s}^{\dagger}
{\hat{c}}_{n s}+{\hat{c}}_{n s}^{\dagger} {\hat{c}}_{n+1 s}\,\right)\nonumber \\ &&
+U\sum_n\,{\hat{c}}_{n\uparrow}^{\dagger}{\hat{c}}_{n \downarrow}^{\dagger} {\hat{c}}_{n
\downarrow}^{} {\hat{c}}_{n\uparrow}^{}\,.
\label{eq:Hel}
\end{eqnarray}
Instead of the Coulomb interaction also between distant lattice sites, only the single-site Hubbard repulsion
  $U_{n,m}=U\delta_{n,m}$ is considered.
The quantities 
$J_{n,n+1}$ denote the transfer matrix elements whose value are 
determined by an overlap integral being responsible for the
nearest-neighbor transport of the electron along the chain.
These transfer matrix elements  are supposed here to depend on the relative distance
between two consecutive units 
in the following
exponential fashion \cite{Slater}
\begin{equation}
J_{n,n+1}=J_0 \,\exp[-\eta\,(u_{n+1}-u_n)]\,.\label{eq:Vexp}
\end{equation}
The quantity $\eta$ regulates how strong the $J_{n,n+1}$ are 
influenced by the relative displacement of lattice units,
or in other words
it determines the coupling strength between the electron and the
lattice system.
The numerical methods to solve the Hubbard model system coupled to the Morse lattice are given in Refs.  \cite{22,37}. 

Compared with the continuum model used in this work, the Coulomb repulsion between the electrons is described in the Hubbard model only in quite \textit{ad-hoc} albeit appropriate  approximation.
We use the comparison with the Hubbard model to control the approximations performed in our approach. In particular, the transition to the continuum limit is possible when the relevant properties vary smoothly in the lattice. With respect to the Coulomb interaction this means that the characteristic length scale given here by the width of the bisolectron $1/\kappa$ is large compared with the lattice parameter. In the opposite case, the simple Coulomb interaction that is singular at $x=0$ cannot be used. A Hubbard Hamiltonian, in general with two-center Coulomb repulsion terms, may become more appropriate.

\begin{figure}[t]
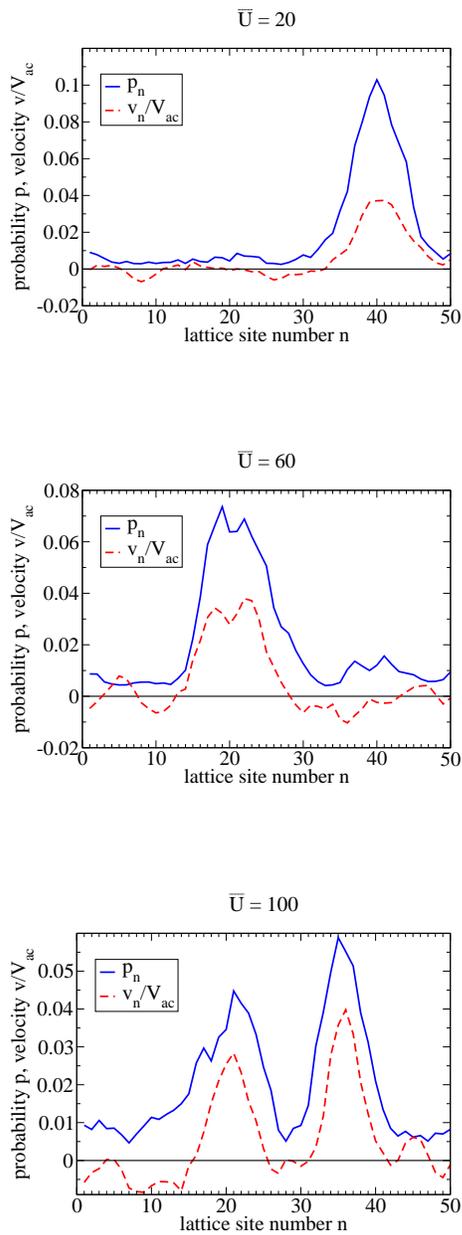

\begin{center}
\includegraphics[width=6cm]{U20.eps}\\
\vspace{1.4cm}
\includegraphics[width=6cm]{U60.eps}\\
\vspace{1.4cm}
\includegraphics[width=6cm]{U100.eps}
\end{center}
\caption{\label{fig5a}
Numerical estimates for the electron density (blue solid line) and the velocity distribution (red dashed line) of solectron pairs with Hubbard repulsion at $\bar{U}=20,60, 100$ on the Morse lattice.}
\end{figure}

The numerical results for electron pairs with single-site Hubbard repulsion 
are shown in
Fig. \ref{fig5a} for the electron pair density (blue solid line) and the particle
velocity distribution (red dashed line). The parameter values used in the simulations were
$\eta = 2.5 a, J_0 = 0.04 \,D$. The adiabaticity parameter $\tau=J_0/ (\hbar \,\Omega_{\rm Morse})$
was fixed at the value $\tau = 20$.

\begin{figure}[t]
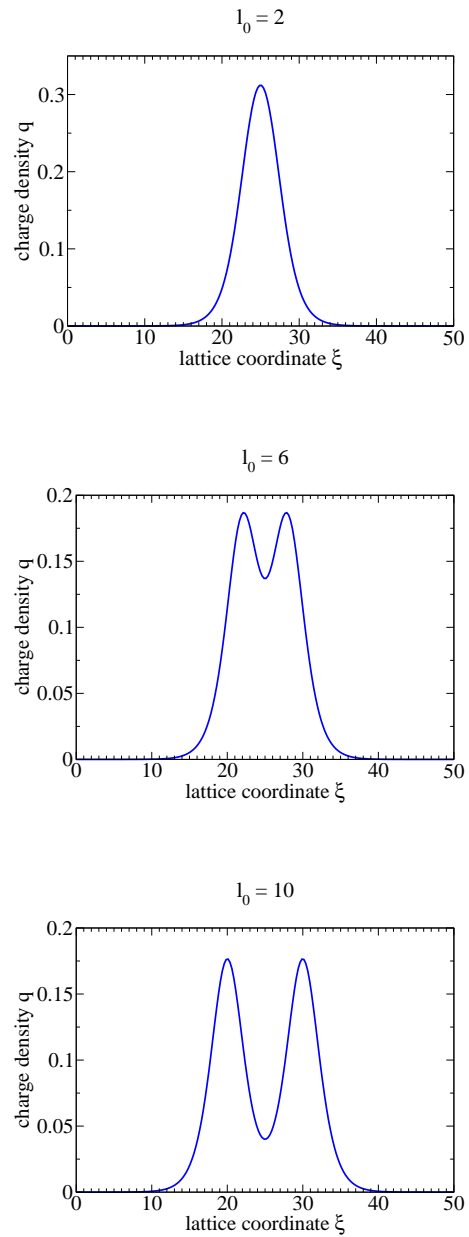

\begin{center}
\includegraphics[width=6cm]{l2.eps}\\
\vspace{1cm}
\includegraphics[width=6cm]{l6.eps}\\
\vspace{1cm}
\includegraphics[width=6cm]{l10.eps}
\end{center}
\caption{\label{fig5b}
Estimates of the bisolectron charge density profile $q(\xi)$, vs. the lattice coordinate $\xi$ using the analytical result (\ref{comsol}) with $\gamma=1.5$, $\kappa = 0.35$,  and $l_0=2,6,10$.}
\end{figure}

We discuss the parameter values that are in the typical range to describe biomolecules more in detail.
The Hubbard parameter was also measured in units of the oscillation energy of the Morse potential using the notation
$\bar{U}=U /\hbar \Omega_{\rm Morse}$. We consider in the numerical simulations the values ${\bar U} = 20, 60, 100$.
We used $\hbar \Omega_{\rm Morse} = 0.002 D$. With values of the depth of the well in the range of
$D \simeq 0.1 - 0.5$ eV we see that the lowest ${\bar U} = 20$ would correspond to $U = 0.004 - 0.02$ eV and the upper value
${\bar U} = 100$ would correspond to $U = 0.02 - 0.1$ eV. Coulomb repulsion energies in this range is what one expects from
physical estimates for electron pairs in a medium with a relative dielectric constant around $\epsilon_{\rm rel} \simeq 10$.
Summarizing we may conclude that Hubbard parameters in the range ${\bar U} = 20 - 100$ which we considered here
correspond to Coulomb repulsion energies in the range $U \simeq 0.01 - 0.05$ eV what seems to make sense.

We see, that depending on the repulsion there can be one or two peaks in the both characteristics, the probability distribution and the velocity distribution. The probability distribution gives the probability  to find an electron at the lattice point under consideration, independent on the position of the other electron. It measures the electron density. The velocity distribution proofs that there is a collective flow as expected for a solitonic excitation. We emphasize that the simulation also shows the stability of the bisolectron that stays bound during the whole run of simulation.

In Fig. \ref{fig5b} we show the charge density function within our analytical model developed in Section III, for various values of the Coulomb repulsion, which determines the distance between the maxima of one-electron functions.  Note that the charge density is determined by (\ref{comsol2}) with $l=l_0$ (\ref{lnull}). We set the ratios of $l_0$ in the three figures the same as the ratios of $\tilde U$ in Fig. \ref{fig5a}.

We cannot expect a one-to-one correspondence between the the numerical and analytical results. They are obtained in different models to treat the anharmonic lattice and the Coulomb repulsion. However, we see a qualitative agreement in both approaches.
In particular, we see that electrons are localized in the bisolectron state, the profile of which depends on the strength of the Coulomb repulsion. The tendency that one maximum splits into two maxima with increasing Coulomb repulsion is clearly seen.

Analyzing the results shown in  Figs. \ref{fig5a}, \ref{fig5b} we see that increasing the Hubbard parameter $U$ has a similar effect as increasing the
distance parameter in our analytical theory that means the increase of the Coulomb repulsin, see Eq. (\ref{lnull1}). Both effects push the centers of the wave functions from another.
Notice, the parameter values used in the numerical simulations, correspond to relatively high non-adiabaticity of the system and strong anharmonicity. Nevertheless, comparison of the three sets of figures corresponding to three different values of the Hubbard term in numerical simulations and Coulomb term in the analytical model shows that our analytical model gives rather good results even for rather strong electron repulsion. The approximations performed in deriving our analytical model, in particular the transition to the continuum description and the expansions in deriving the nonlinear potential and the separation of the center of mass motion from the relative motion, seem not to be decisive for the qualitative behavior of the bisolectron solution in the parameter range considered here. 

\section{Conclusions}

We have shown that the anharmonicity of a lattice favors self-trapping and electron pairing in the lattice soliton deformation well. In the particular case of a cubic anharmonicity, the explicit expressions for the electron wave function and the traveling deformation of soliton type are obtained here.We give here also the first excited state which is a triplet state of two electrons with parallel spins with antisymmetric spatial wave function. We find also  limits of the continuum description  if the relative distance between the electrons becomes comparable with the lattice parameter for strongly localized electrons. Then, the singular Coulomb interaction ($\propto 1/|x|$) should be replaced by the matrix elements with respect to the tight-binding orbitals and higher order derivatives in the continuum approximation should be taken into account.



The anharmonic bisoliton, found here and denoted bisolectron (a soliton binding two electrons with Coulomb repulsion and Pauli's principle incorporated), can move with velocity up to the sound velocity, with both energy and momentum, maintaining finite values also at the sound velocity. Low-dimensional systems, mentioned in the Introduction, are characterized by the parameter values, for which the adiabatic approximation is valid. 
Their ground electron state is soliton-like and is extended over a few lattice sites. Therefore, we expect that in these systems pairing of electrons takes place at enough level of doping and that such bisolectron states are stable with account of the Coulomb repulsion and at the velocities up to the velocity of sound. Typical values for poly-peptide macromolecules have been considered that allow for the formation of a bisolectron bound state.

Comparison of the energy of such a bisolectron with the energy of the two independent solitons binding a single electron each (two separate solectrons) shows that there is the gain of energy even including the Coulomb repulsion. 
For illustration we have estimated the bisolectron binding energy for parameter values typical for biological macromolecules. We found that the lattice anharmonicity significantly enhances pairing of electrons. 
We discuss the wave function of relative motion of two electrons in detail and include triplet states which lead to finite values of the Coulomb energy.  Noteworthy is that there is a rather good qualitative agreement of the analytical model for two electrons in a singlet bisolectron state with account of Coulomb repulsion, developed here and the numerical simulations for the Hubbard model.

%

\begin{acknowledgments}

The authors are grateful to A.S. Alexandrov, E. Br\"and\"as, L. Cruzeiro, F. de Moura, J. Feder, R. Lima, G. Vinogradov and E.G. Wilson for enlightening discussions and/or correspondence. This research was supported by the Spanish Ministerio de Ciencia e Innovaci{o}n under grants EXPLORA FIS2009-06585 and MAT2011-26221. L. Brizhik acknowledges partial support from the Fundamental Research Grant of the National Academy of Sciences of Ukraine.
\end{acknowledgments}



\begin{thebibliography}{99}


\bibitem{polydiacet1}
E. G.  Wilson, J. Phys. C: Solid State Phys. {\bf 16},  6739 (1983).

\bibitem{polydiacet2}
K. J. Donovan, and E. G. Wilson, Phil. Mag. B {\bf 44}, 9, 31 (1981).

\bibitem{Donovan}
K. J. Donovan, and E. G. Wilson, J. Phys. C {\bf 18}, L-51 (1985).

\bibitem{polydiacet3}
A. S. Gogolin, JETP Lett. {\bf 43}, 511 (1986).

\bibitem{Monceau}
P. Monceau (Ed.), {\it Electronic  Properties of Inorganic Quasi-One-Dimensional Compounds}, Part II  (Reidel, Dordrecht, 1985).

\bibitem{trans-met1}
B. G. Streetman, and S. K. Banerjee, {\it  Solid State Electronic Devices}  (Prentice-Hall, N.J., 2009).

\bibitem{trans-met2}
Y. Zhang, X. Ke, C. Chen, and P. C. Kent, Phys. Rev. B \textbf{80}, 024303 (2009).

\bibitem{trans-met3}
O. Madelung, U. R\"ossler, and M. Schultz (Eds.), {\it PdO Crystal Structure, Lattice Parameters, Thermal Expansion.}  V. 41D: {\it Non-Tetrahedrally Bonded elements and Binary Compounds I} (Springer,  Berlin, 1998).

\bibitem{trans-met4}
J. Androulakis, Y. Lee, I. Todorov, D.-Y. Chung, and M. Kanatzidis, 
Phys. Rev. B {\bf 83} 195209 (2011).

\bibitem{cuprates1}
C. Falter, and G. A. Hoffmann, Phys. Rev. B {\bf 64}, 054516 (2001).

\bibitem{cuprates2}
K.-P. Bohnen, R. Heid, and M. Krauss, Europhys. Lett. {\bf 64}, 104 (2003).

\bibitem{cuprates3}
T. P. Devereaux, T. Cuk, Z.-X. Shen, and N. Nagaosa, Phys. Rev. Lett. {\bf 93}, 117004 (2004).

\bibitem{cuprates4}
D. Reznik, L. Pintschovius, M. Ito, S. Iikubo, M. Sato, H. Goka, M. Fujita, K. Yamada, G. D. Gu, and J. M. Tranquada, Nature {\bf 440}, 1170 (2006). 

\bibitem{8}
 V. Z. Kresin, and S. A. Wolf, Rev. Mod. Phys. {\bf 81}, 481 (2009). 
 
\bibitem{7}
D. M. Newns, and C. C. Tsuei, Nature Physics {\bf 3}, 184 (2007). 

\bibitem{ASD}
A. S. Davydov, {\it Solitons in Molecular Systems} (Reidel, Dordrecht, 1991).

\bibitem{Scott}
A. C. Scott,  Phys. Rep. \textbf{217}, 1 (1992).

\bibitem{Chris}
P. L. Christiansen, and A. C. Scott (Eds.), {\it Davydov's Soliton Revisited. Self-trapping of Vibrational Energy in Protein} (Plenum, New York, 1990).

\bibitem{Bountis}
T. Bountis (Ed.), {\it Proton Transfer in Hydrogen-Bonded Systems} (Plenum, New York, 1992)

\bibitem{Dauxois}
T. Dauxois, and M. Peyrard, {\it Physics of Solitons} (Cambridge University Press, Cambridge, 2006).

\bibitem{Yakush}
L. V. Yakushevich, \textit{Nonlinear Physics of DNA} (Wiley-VCH, Weinheim, 2004).

\bibitem{Peyrard}
M. Peyrard, and J. Farago,  Physica A \textbf{288}, 199 (2000).


\bibitem{Manevich}
L. I. Manevich, and V. V. Simmons, {\it Solitons in Macromolecular Systems} (Nova, New York, 2008).

\bibitem{16}
M. G. Velarde, J. Computat. Applied Maths. {\bf 233}, 1432  (2010).

\bibitem{Landau}
L. D. Landau, Phys. Z. Sowjetunion {\bf 3}, 664 (1933).

\bibitem{Pekar}
S. I. Pekar {\it Untersuchungen \"uber die Elektronentheorie} (Akademie Verlag, Berlin, 1954).

\bibitem{Kaganov}
M. I. Kaganov, and I. M. Lifshitz, {\it Quasiparticles} (Mir, Moscow, 1979).

\bibitem{5}	
A.S. Alexandrov, and N. Mott, {\it Polarons and Bipolarons} (World Scientific, Cambridge, 1995).

\bibitem{Alex}
A. S. Alexandrov (Ed.), {\it Polarons in Advanced Materials} (Springer, Dordrecht, 2007).

\bibitem{Alex2}
A. S. Alexandrov, and J. T. Devreese, {\it Advances in Polaron Physics} (Springer, Dordrecht, 2009).

\bibitem{6}
J. T. Devreese (Ed.), {\it Polarons in Ionic Crystals and Polar Semiconductors} (North-Holland, Amsterdam 1972).

\bibitem{ASD-Zolot1}
A. S. Davydov, and A. V. Zolotaryuk, Phys. Lett.  {\bf 94}, 49  (1983).

\bibitem{ASD-Zolot3}
A. S. Davydov, and A. V. Zolotaryuk,  Phys. Stat. Sol. (b) {\bf 115},  115 (1983).

\bibitem{ASD-Zolot2}
A. S. Davydov, and A. V. Zolotaryuk, Physica Scripta {\bf 30}, 426 (1984).

\bibitem{Briz-Dav}
L. S. Brizhik, and A. S. Davydov, 
J. Low Temp. Phys. {\bf 10}, 748  (1984).

\bibitem{13}
L. S. Brizhik,  J. Low Temp. Phys. {\bf 12}, 437 (1986).

\bibitem{Briz-Er}
L. S. Brizhik, and A. A. Eremko, 
Physica D {\bf 81},  295 (1995).

 \bibitem{1}
 M. G. Velarde, L. Brizhik, A. P. Chetverikov, L. Cruzeiro, W. Ebeling, and G. R\"opke, Int. J. Quantum Chem. {\bf 112}, 551 (2011).

  \bibitem{2}
M. G. Velarde, L. Brizhik, A. P. Chetverikov, L. Cruzeiro, W. Ebeling, and G. R\"opke, Int. J. Quantum Chem. D.O.I. 10.1002/qua.23282 (2012) (to appear).

\bibitem{3}
J. Hubbard, Proc. Roy. Soc. (London) A {\bf 276}, 238 (1963); A {\bf 277}, 237 (1963); A {\bf 281}, 401 (1964).

\bibitem{4}
A. Montorsi (Ed.), {\it The Hubbard Model. A Reprint Volume} (World scientific, Singapore 1992).

\bibitem{16a}
A. V. Zolotaryuk, K. H. Spatschek, and A. V. Savin, Phys. Rev. B {\bf 54}, 266 (1996).

\bibitem{20}
D. Hennig,  M. G. Velarde, W. Ebeling, and A. P. Chetverikov, Phys. Rev. E {\bf 78}, 066606 (2008).


\bibitem{23}
W. Ebeling, M. G. Velarde, A. P. Chetverikov, and D. Henning,
in {\it Selforganization of Molecular Systems. From Molecules and Clusters to Nanotubes and Proteins}, N. Russo, V.Ya. Antonchenko, E. Kryachko (Eds.) (Springer, Berlin 2009) pp. 171-198.

\bibitem{23a}
A. P. Chetverikov, W. Ebeling, M. G. Velarde, Eur. Phys. J. B {\bf 70}, 217 (2009).


\bibitem{22}
M. G. Velarde, W. Ebeling, and A. P. Chetverikov, Int. J. Bifurc. Chaos {\bf 21}, 1595 (2011).

\bibitem{Neissner}
M. G. Velarde, and C. Neissner, Int. J. Bifurcation Chaos {\bf 18}, 885 (2008).

\bibitem{18}
L. Cruzeiro, J. C. Eilbeck, J. L. Marin, and F. M. Russell, Eur. Phys. J. B {\bf 42}, 95 (2004).

\bibitem{dias}
W. S. Dias, M. L. Lyra, and F. A. B. F. de Moura, Phys. Lett. A {\bf 374}, 4554 (2010).

\bibitem{dias2}
W. S. Dias, E. M. Nascimento, M. L. Lyra, and F. A. B. F. de Moura, Phys. Rev. B {\bf 81}, 045116 (2010).

\bibitem{dias3}
W. S. Dias, M. L. Lyra, and F. A. B. F. de Moura, Eur. Phys. J. B {\bf 85}, 7 (2012).


\bibitem{30}
M. Toda, {\it Theory of Nonlinear Lattices}, 2nd ed. (Springer, Berlin, 1989).

\bibitem{31}
J. Dancz, and S. A. Rice, J. Chem. Phys. {\bf 67}, 1418  (1977).

\bibitem{32}
T. J. Rolfe, S. A. Rice, and J. Dancz, J. Chem. Phys. {\bf 79}, 26  (1979).

\bibitem{Slater}
J. C. Slater, {\it Quantum Theory of Molecules and Solids}, vol. 4 (McGraw-Hill, New York, 1974).

\bibitem{37}
W. Ebeling, M. G. Velarde, and A. P. Chetverikov,  Cond. Matter Phys. {\bf 12}, 633 (2009).  























\end{thebibliography}
\end{document}